\documentclass[sigconf]{nonacmart}

\usepackage{courier}
\usepackage{graphicx}
\usepackage{graphics}
\usepackage{url}
\usepackage{multirow}
\usepackage{color}
\usepackage{subfig}
\usepackage{balance}

\renewcommand\footnotetextcopyrightpermission[1]{} 		

\newcommand{\superscript}[1]{\ensuremath{^{#1}}}

\def\iit{\superscript{*}}
\def\mpi{\superscript{\#}}

\begin{document}

\clubpenalty = 10000
\widowpenalty = 10000

\setlength{\belowdisplayskip}{1pt} 
\setlength{\belowdisplayshortskip}{1pt}
\setlength{\abovedisplayskip}{1pt} 
\setlength{\abovedisplayshortskip}{1pt}

\title{On Quantifying Knowledge Segregation in Society}


\author{
  Abhijnan Chakraborty\mpi\iit,~~~~~~~~
  Muhammad Ali\mpi,~~~~~~~~
  Saptarshi Ghosh\iit,~\\
  Niloy Ganguly\iit,~~~~~~~~~~
  Krishna P. Gummadi\mpi \\
  ~\\
  {\mpi}Max Planck Institute for Software Systems, Germany \\
  {\iit}Indian Institute of Technology Kharagpur, India
}

\begin{abstract}
With rapid increase in online information consumption, especially via social media sites, there have been concerns on whether people are getting {\it selective exposure} to a biased subset of the information space, where a user is receiving more of what she already knows, and thereby potentially getting trapped in {\it echo chambers} or {\it filter bubbles}. Even though such concerns are being debated for some time, it is not clear how to quantify such echo chamber effect. In this position paper, we introduce {\it Information Segregation} (or {\it Informational Segregation}) measures, which follow the long lines of work on residential segregation. We believe that information segregation nicely captures the notion of exposure to different information by different population in a society, and would help in quantifying the extent of social media sites offering selective (or diverse) information to their users.
\end{abstract}

\maketitle

\section{Introduction}
As increasing number of users are consuming information online, often via social media sites like Facebook and Twitter, there have been concerns regarding 
the content quality~\cite{agichtein2008finding,chakraborty2016stop}, and 
the possibility of {\it biases} in the information people are getting exposed to~\cite{colleoni2014echo,chakraborty2015can,chakraborty2016dissemination,chakraborty2017makes}. In such sites, people tend to be connected with other like-minded users out of homophily~\cite{mcpherson2001birds,aiello2012friendship}, and create their own interest groups~\cite{bhattacharya2014deep,chakraborty2013clustering}.   
Thus, there have been concerns that individual users can have {\it selective exposure} to information which closely matches their own views, 
and may not have enough exposure to differing views~\cite{colleoni2014echo}, and such {\it  echo chambers} or {\it filter bubbles}~\cite{pariser2011filter} 
may lead to the {\it polarization} of society~\cite{dandekar2013biased,mas2013differentiation}.

However, two competing theories of opinion polarization have been proposed in earlier works~\cite{maes2015will}. One school of thought assumes that opinions are reinforced when likeminded individuals interact with each other~\cite{dandekar2013biased,mas2013differentiation,friedkin2011social}. 
 Whereas, other researchers have argued that exposure to differing views and their subsequent rejections lead to polarization~\cite{baldassarri2008partisans,mason2007situating}. 
 Polarization can be thought as a {\it measure of the ideological state} of the population in a society, which is difficult to quantify in general. Also, it is not explicitly clear what constitutes the ideal notion of the {\it depolarized} state of a society.
 
 \begin{figure}[htb]
\center{
\vspace{-2mm}
\includegraphics[width=0.7\columnwidth]{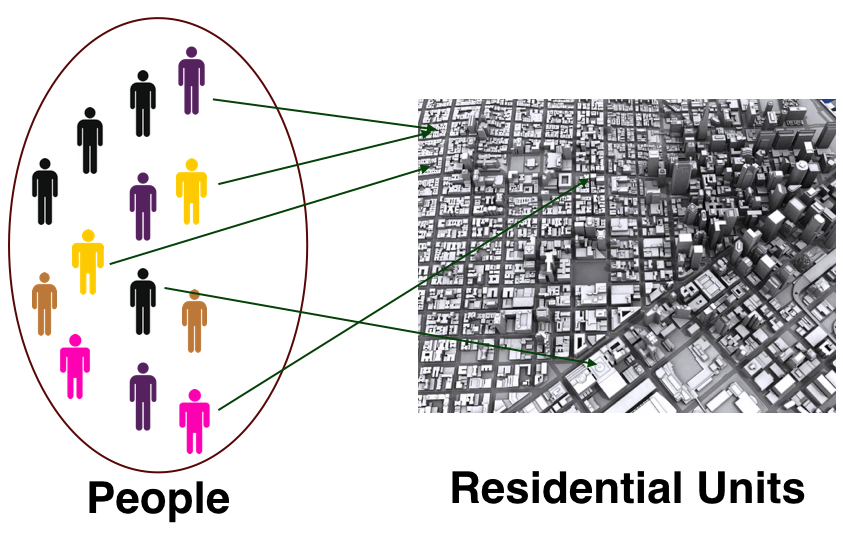}
}
\vspace{-2mm}
\caption{{\bf Basis for computing residential segregation: bipartite matching between people and residential units in a city.}}
\vspace{-2mm}
\label{fig:user_residential_mapping}
\end{figure}

In this position paper, we argue that an alternative option would be to consider the access to different types of information by members of a society. 
For example, within a population with multiple parties operating, it is but natural that political 
opinion would be fragmented. 
However, it is highly desirable that the entire population have access to the same information / knowledge and they take informed decision to follow different paths. 
In other words, the bigger issue here is {\it whether different groups of people are having access to similar kind of information or not}, where groups may be formed based on predefined demographics (e.g., gender, race, age, income level) or derived features (e.g., political leaning) of people. 

 \begin{figure}[htb]
\center{
\vspace{-2mm}
\includegraphics[width=0.7\columnwidth]{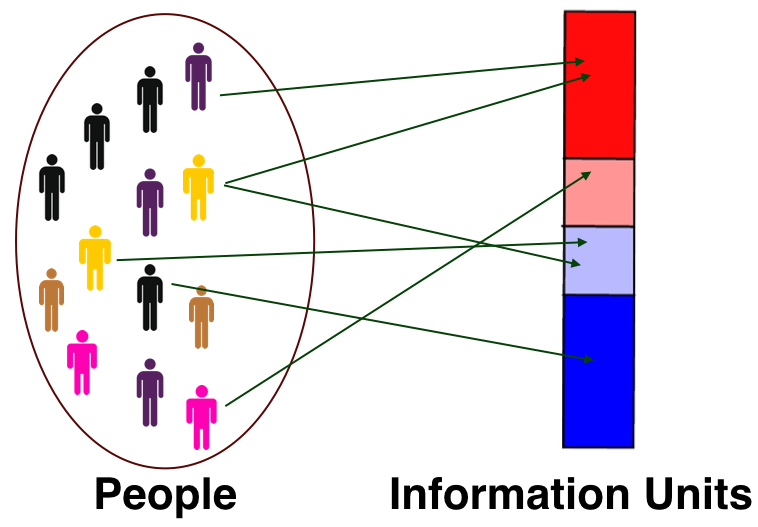}
}
\vspace{-2mm}
\caption{{\bf Basis for computing information segregation: bipartite matching between people and information units.}}
\vspace{-2mm}
\label{fig:user_information_mapping}
\end{figure}

To investigate this issue, we borrow ideas from the past literature on {\it residential segregation}.
A large number of research works have considered the bipartite matching between different groups of people 
and the urban units where they reside (as shown in Figure~\ref{fig:user_residential_mapping}),  
and proposed different measures to quantify geographical segregation of different groups~\cite{dorfman1979formula,duncan1955residential}. 
In a seminal work, 
Massey and Denton~\cite{massey1993american} identified five distinct dimensions of residential segregation: \\ 
(i)   {\bf Evenness} is the degree to which groups are distributed proportionately across areal units in an urban area. \\
(ii)  {\bf Exposure} is the extent to which members of different groups share common residential areas. \\
(iii) {\bf Concentration} refers to the degree of a group's agglomeration in urban space. \\
(iv) {\bf Centralization} is the extent to which group members reside towards the center of an urban area, and \\
(v)  {\bf Clustering} measures the degree to which different groups are located adjacent to one another.

Then, they grouped different segregation measures along these five dimensions. Note that some segregation measures are {\it relative between two groups}, whereas others are {\it absolute measures of the segregation of one particular group}.

Following this line of work, in this paper, we present the notion of {\bf Information Segregation} (or {\bf Informational Segregation}). Similar to Figure~\ref{fig:user_residential_mapping}, we consider another bipartite matching between different groups of people and the information units they have access to (shown in Figure~\ref{fig:user_information_mapping}). Then utilizing this mapping, we can compute information segregation to measure whether different groups in a society are having access to similar kind of information or not.

However, there are two primary aspects where the mapping between people and information units differs from the mapping between people and residential units: 
(i)~residential segregation is computed over a two-dimensional geographical space, whereas information segregation needs to be computed over a $n$-dimensional topic space ($n = 1$ in Figure~\ref{fig:user_information_mapping}, but in general, $n \ge 1$), and 
(ii)~one person may have access to multiple information units, which needs to be accounted for while computing information segregation; whereas, 
one person is considered to be permanently staying in only one residential unit. 
To account for people accessing different information units, we use the notion of {\bf fractional personhood}~\cite{perring1997degrees}. For an information unit $i$, we consider the personhood of $1$ for everyone who have access to only $i$, 
personhood of $\frac{1}{2}$ for them who have access to $i$ and another information unit, and so on.  

In this paper, we propose five measures of information segregation analogous to the residential segregation measures discussed earlier, by considering the fractional personhoods of people from different groups. Then, as a proof of concept, we measure the information segregation of US-based Facebook users as evident from how they follow different news media pages on Facebook. Our investigation reveals that Hispanic users are accessing information more evenly across political spectrum; whereas Asian Americans have highest information segregation among all racial groups. Similarly, we also looked at how users having different political leanings are accessing contrary views. We found that moderately conservative leaning users tend to get information more evenly across the spectrum; whereas, extremely conservative leaning users are most segregated among others.

The information segregation measures proposed in this paper can also be used to evaluate the role of search / recommender systems for exposing different types of information to a large population. We believe that in future, greater emphasis should be put on designing more responsible search / recommender systems which limit information segregation to acceptable limits. 

\section{Measures of Information Segregation}
In this section, we introduce different measures of information segregation, considering the 
five distinct dimensions as identified by Messey and Denton~\cite{massey1993american} for residential segregation. 

\begin{figure}[htb]
\center{
\vspace{-3mm}
\includegraphics[width=0.7\columnwidth]{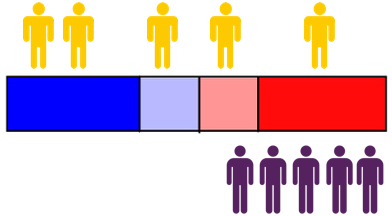}
}
\vspace{-2mm}
\caption{{\bf Yellow group gets information more evenly than Purple group.}}
\vspace{-3mm}
\label{fig:evenness}
\end{figure}

\vspace{1mm}
\noindent {\bf I. Evenness} \\
{\it The evenness measure of information segregation captures how uniformly members of a particular group have access to   
different units in the $n$-dimensional information space.} Figure~\ref{fig:evenness} shows an example scenario where members of Yellow group have access to all four information units; whereas, members of Purple group have access to only two units. Therefore, Yellow group in Figure~\ref{fig:evenness} have more even information access than Purple group. 
Massey and Denton~\cite{massey1993american} discussed five different measures of residential evenness (including both relative and absolute measures). 
For 
brevity, we are defining only one measure of absolute evenness of a group, 
which is the complement of {\bf Gini Coefficient}~\cite{dorfman1979formula}. 

Gini coefficient $G_{A}$ measures the unevenness of a particular group $A$, by capturing the mean absolute difference between the 
personhoods of $A$ having access to different information units. Then, Information Evenness $IE_A$ can be computed as 
$$
IE_A = 1 - G_{A} = 1 - \frac{\sum_{i = 1}^{m} \sum_{j=1, j \neq i}^{m} |a_i - a_j|}{2 \cdot a_{total} \cdot a_{total}^{'}} 
$$
where $a_i$ is the sum of personhoods belonging to group $A$ who get information $i$, $a_{total}$ is the size of group $A$ in the overall population, $m$ is the number of information units, and $a_{total}^{'}$ is the number of people in the overall population who {\it do not belong} to group $A$. $IE_A$ varies between $0$ to $1$, higher the value, the group has more even information access. 

\begin{figure}[htb]
\center{
\vspace{-3mm}
\includegraphics[width=0.7\columnwidth]{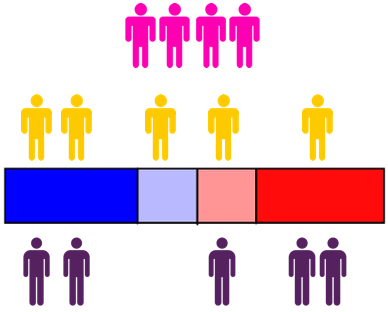}
}
\vspace{-2mm}
\caption{{\bf Joint exposure between Purple group and Yellow group is higher than the joint exposure between Purple group and Pink group.}}
\vspace{-3mm}
\label{fig:exposure}
\end{figure}

\vspace{1mm}
\noindent {\bf II. Joint Exposure} \\
{\it Joint exposure quantifies the extent to which members of two groups get jointly exposed to the same information.}
In Figure~\ref{fig:exposure}, members of Purple and Yellow groups are jointly exposed to three out of four information units; whereas, members of Purple and Pink groups are jointly exposed to only one unit. Therefore, in Figure~\ref{fig:exposure}, Purple and Yellow groups have higher joint exposure compared to Purple and Pink groups.
 
Again using the notion of personhoods, joint information exposure between groups $A$ and $B$ is computed as 
$$
JIE_{AB} = \sum_{i = 1}^{m} \frac{a_i}{a_{total}} \cdot \frac{b_i}{total_i} 
$$
where $a_i$, $a_{total}$, and $m$ are as defined earlier, $b_i$ is sum of personhoods belonging to $B$ who get information $i$, and $total_{i}$ is sum of all personhoods having access to information $i$. $JIE_{AB}$ varies between $0$ to $1$, higher the value, $A$ and $B$ have more common exposure.

\begin{figure}[t]
\center{
\vspace{-3mm}
\includegraphics[width=0.7\columnwidth]{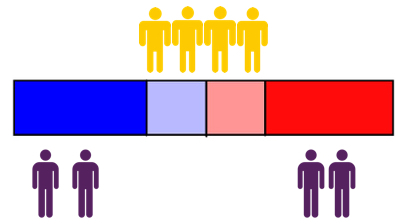}
}
\vspace{-2mm}
\caption{{\bf Yellow group is more concentrated than Purple group.}}
\vspace{-3mm}
\label{fig:concentration}
\end{figure}

\vspace{1mm}
\noindent {\bf III. Concentration} \\
{\it Concentration of a group $A$ refers to the relative amount of topical space that $A$ have access to.} 
Every information unit may not have similar topical density (or number of information sources, etc), 
with some units having more topics mapping into it, compared to other information units. For example, in Figure~\ref{fig:concentration}, 
red and blue units consist of higher number of topics than blueish and reddish grey units. 
Therefore, even though Yellow and Purple groups have access to same number of units (hence have same evenness), Yellow group  
 would be considered more concentrated (i.e., more segregated) as it has access to fewer topics.
Information concentration is captured by the metric {\bf Delta}~\cite{hoover1941interstate}:
$$
DEL_{A} = \frac{1}{2} \sum_{i = 1}^{m} \frac{a_i}{a_{total}} \cdot \frac{n_i}{n_{total}}
$$
where $a_i$, $a_{total}$, and $m$ are already defined, $n_i$ is number of topics in information unit $i$, and $n_{total}$ is number of topics overall.

\begin{figure}[htb]
\center{
\vspace{-3mm}
\includegraphics[width=0.7\columnwidth]{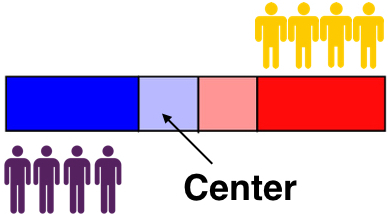}
}
\vspace{-2mm}
\caption{{\bf Purple group is more centralized than Yellow group.}}
\vspace{-3mm}
\label{fig:centralization}
\end{figure}

\vspace{1mm}
\noindent {\bf IV. Centralization} \\
Compared to the geographical context, identifying the center of an information space is tricky, and may not be always possible.  
Centrality may be computed by considering centroids in a dimension-reduced topical space, 
or by measuring it over networks induced by information units and their topical or preference similarity. 
In scenarios where the notion of information center is defined, {\it centralization between two groups $A$ and $B$ refers to how the information units that $A$ and $B$ have access to are distributed around the center.} For example, in Figure~\ref{fig:centralization}, if we assume the blueish grey unit to be the center, then although Yellow and Purple groups have same evenness and concentration measures, Purple group is more centralized than Yellow group.
Formally, {\bf Centralization Index}~\cite{duncan1955residential} can be measured as 
$$
CI_{AB} = \sum_{i = 1}^{m} a_{i-1}{b_{i}} - \sum_{i = 1}^{m} a_{i}{b_{i-1}}
$$
where information units are sorted based on their distance from the center, and $a_i$, $b_i$, and $m$ are as defined earlier. $CI_{AB}$ varies between $-1$ to $1$, positive value indicating $A$ is more centralized than $B$.

\begin{figure}[t]
\center{
\vspace{-3mm}
\includegraphics[width=0.7\columnwidth]{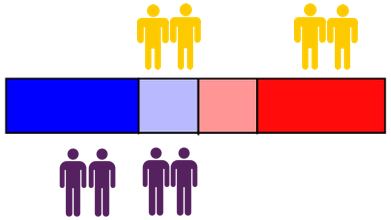}
}
\vspace{-2mm}
\caption{{\bf Purple group is more clustered than Yellow group.}}
\vspace{-3mm}
\label{fig:clustering}
\end{figure}

\vspace{1mm}
\noindent {\bf V. Clustering} \\
The final dimension of information segregation is the degree to which members of a group $A$ have access to information clusters, i.e., {\it whether the different types of information received by $A$ are close to each other in the information space.} 
In Figure~\ref{fig:clustering}, both Purple and Yellow groups have access to two information units, and have the same evenness and concentration scores. However, as the information units Purple group have access to are close to each other, according to clustering measure, it is more segregated than Yellow group .
We can formally define information clustering as
$$
IC_{A} = \frac{(\sum_{i = 1}^{m} \frac{a_i}{a_{total}} \sum_{j = 1}^{m} e^{-d_{ij}} a_j) - (\frac{a_{total}}{m^2} \sum_{i = 1}^{m} \sum_{j = 1}^{m} e^{-d_{ij}})}{(\sum_{i = 1}^{m} \frac{a_i}{a_{total}} \sum_{j = 1}^{m} e^{-d_{ij}} total_j) - (\frac{a_{total}}{m^2} \sum_{i = 1}^{m} \sum_{j = 1}^{m} e^{-d_{ij}})}
$$
where $a_i$, $a_{total}$, $total_j$, and $m$ are as defined earlier, and $d_{ij}$ is the distance between information units $i$ and $j$. $IC_{A}$ varies from $0$ to $1$.

\begin{figure*}[tb]
\vspace{-2mm}
\center{
\subfloat[{\bf }]{\includegraphics[width=0.28\textwidth]{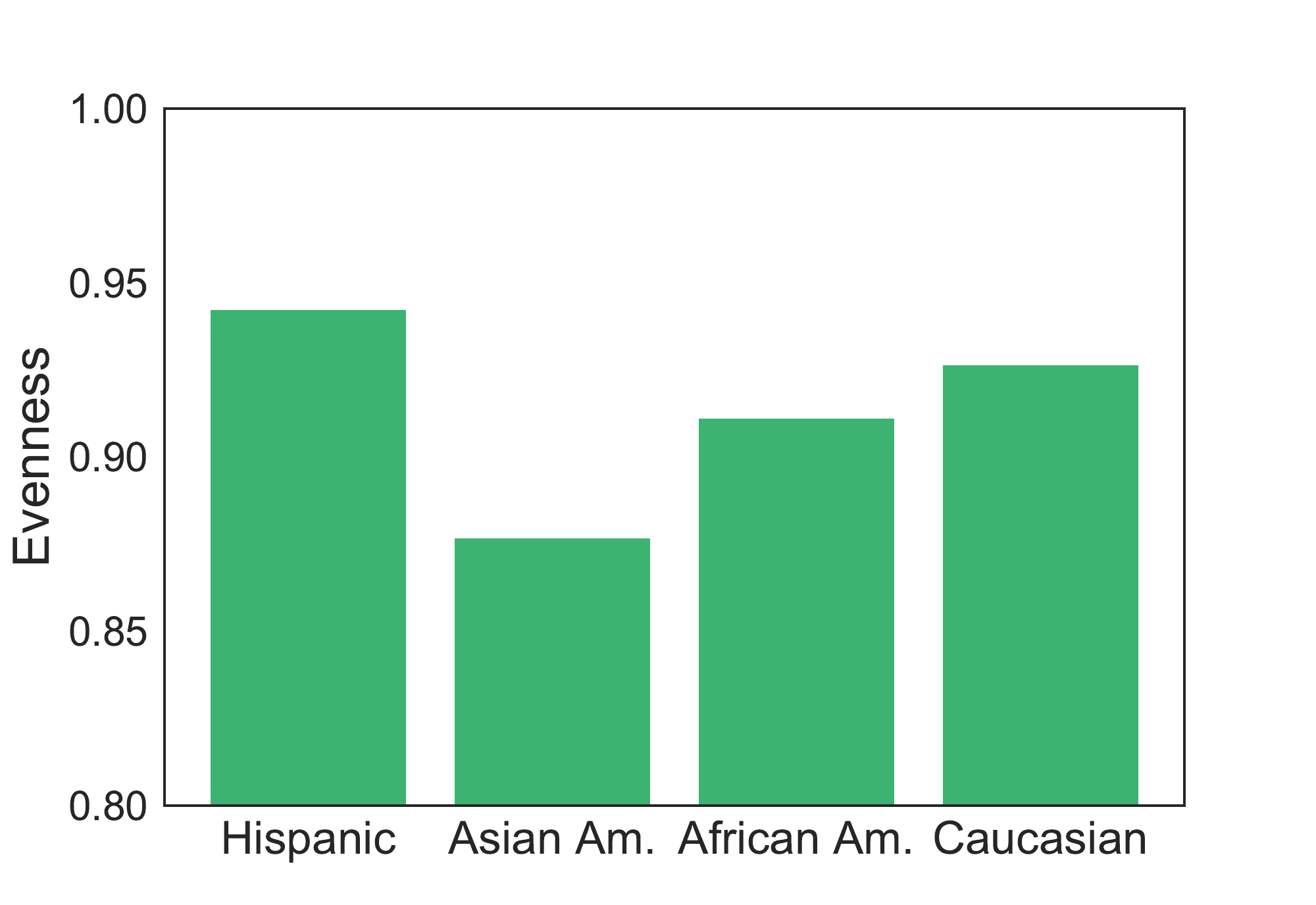}}
\hfil
\subfloat[{\bf }]{\includegraphics[width=0.28\textwidth]{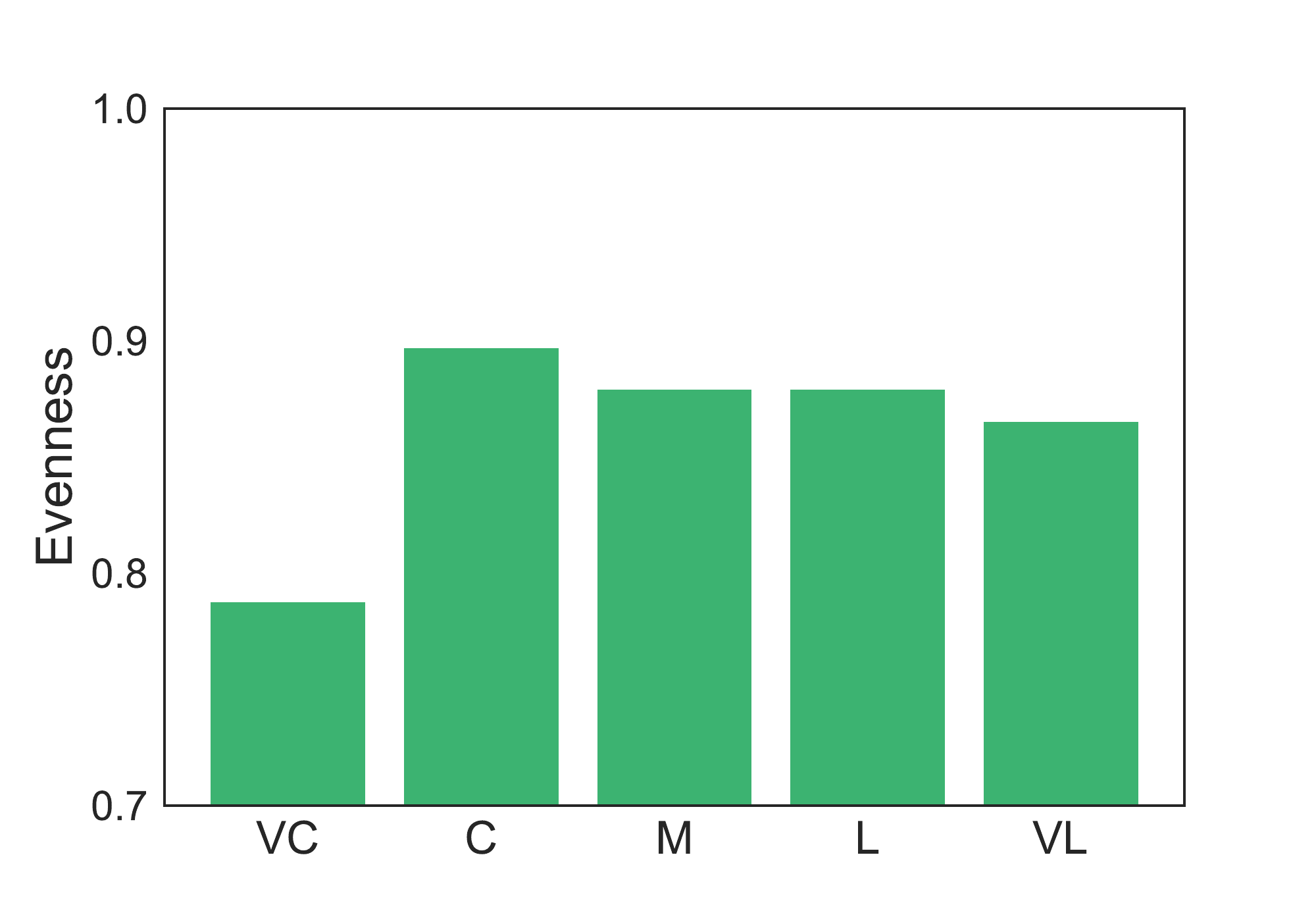}}
\hfil
\subfloat[{\bf }]{\includegraphics[width=0.28\textwidth]{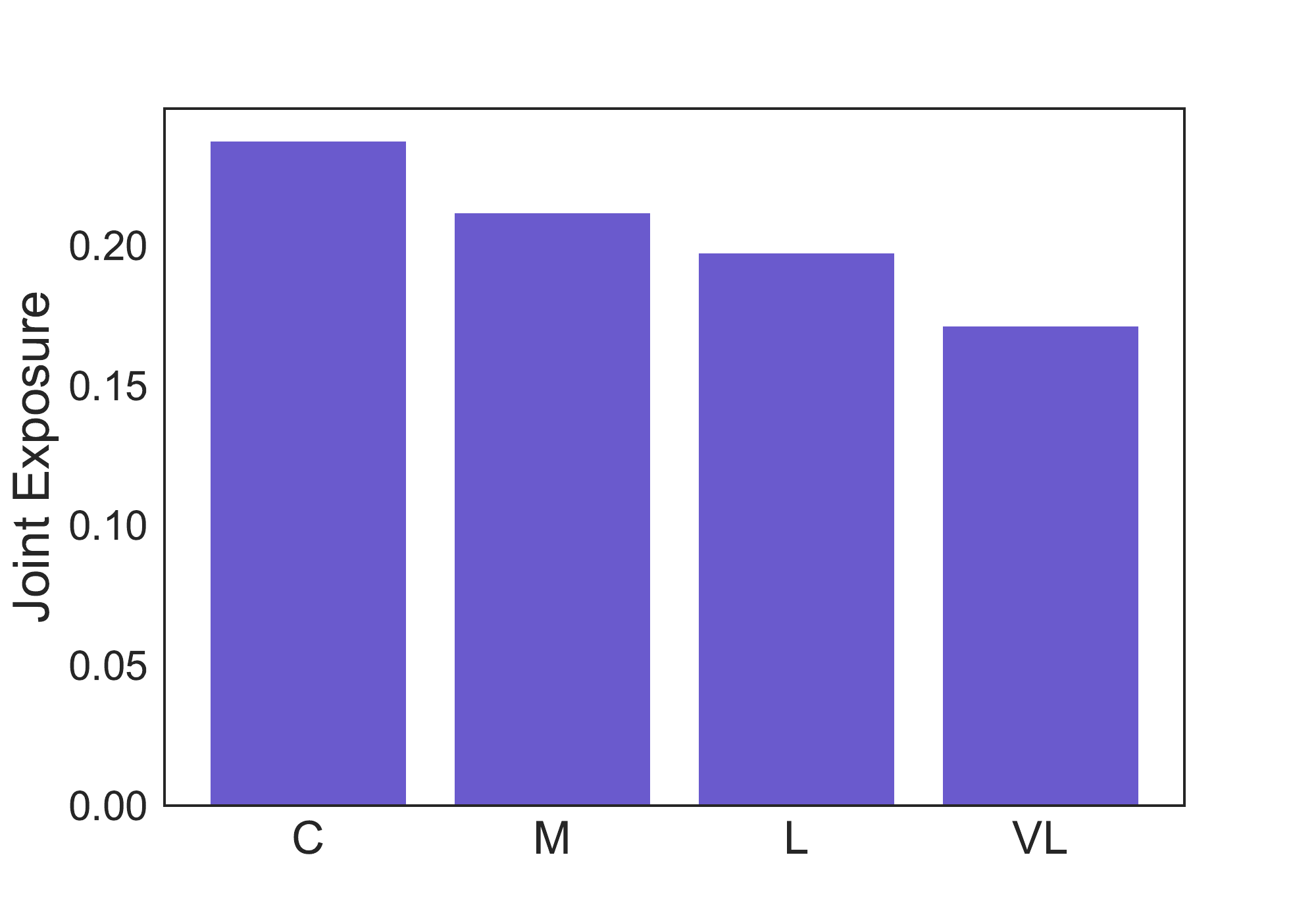}}
}
\vspace{-2mm}
\caption{{\bf Information segregation between different groups along two dimensions: evenness of 
(a)~different racial groups, (b)~different political groups, and (c)~joint exposure of very conservative leaning people (VC) with other political groups.}}
\vspace{-3mm}
\label{fig:segregation}
\end{figure*}

\section{Information Segregation among US-Based Facebook Users} 
Next, we attempt to quantify information segregation of Facebook users in the US. Towards that end, we specifically focus on news media pages in Facebook, and measure information segregation with respect to how different groups of users follow these pages. 

\vspace{1mm}
\noindent {\bf Dataset Gathered} \\
We queried Facebook search with the term `US news media' to collect US related news media pages in Facebook, and found more than $2.5K$ Facebook pages for that query. 
Then using Facebook's ad submission web page ({\it facebook.com/ads/manager/creation}), we collected the composition of gender, race and political leanings of the followers of these media pages. We acknowledge the limitation that the retrieved pages may not be representative of all US media pages, and we would expand the corpus in future work.

\vspace{1mm}
\noindent {\bf Mapping Facebook Pages to Information Units} \\
To quantify information segregation, we focus on $1$-dimensional political information space, and divide it into five information units: {\bf Very Conservative (VC), Conservative (C), Moderate (M), Liberal (L),} and {\bf Very Liberal (VL)}. Then, we map different news pages on Facebook to one of these five information units by considering the political leanings of the followers of these pages. 
For a page $P$, if the fraction of followers leaning towards 
respective political ideologies are denoted as $f_{VC}, f_{C}, f_{M}, f_{L}$, and $f_{VL}$ respectively, then we measure the political leaning of P ($Leaning_{P}$) 
as a weighted sum of the political leaning of its audience. More specifically, 
$$
Leaning_{P} = -1 \cdot f_{VC} + -0.5 \cdot f_{VC} + 0 \cdot f_{M} + 0.5 \cdot f_{L} + 1 \cdot f_{VL}
$$ 
If $Leaning_{P}$ is between $-0.1$ to $+0.1$, we map $P$ to information unit $M$; for $Leaning_{P}$ between $0.1$ to $0.5$, $P$ is mapped to $L$ and for $Leaning_{P} > 0.5$, we map $P$ to $VL$. Similarly, we map $P$ to $C$ or $VC$ if $-0.5 \le Leaning_{P} < -0.1$ and $Leaning_{P} < -0.5$ respectively.

\vspace{1mm}
\noindent {\bf Computing the Personhood Scores} \\
After mapping every page to one of the information units, we try to gather the cumulative number of followers 
for a particular unit. However, Facebook doesn't allow us to get the follower size for a combination of more than $400$ Facebook pages. Therefore, we randomly select $400$ pages from the set of $2.5K+$ news media pages, map them to their corresponding units, and gather the demographics of the followers of pages belonging to every information unit.

As some users may follow Facebook pages belonging to multiple units (for example, follow both conservative and liberal leaning pages), we need to accurately account for these overlaps in information access. As mentioned earlier, we use the notion of {\it fractional personhood} in this regard. Therefore, instead of considering the number of followers of pages in a particular unit, we consider the sum of personhoods for pages in every information unit.

For every unit $i$, the sum of personhoods $N_i^{*}$ is computed as
$$
N_i^{*} = [N(S) - N(S \setminus i)] + \frac{1}{2} \sum_{j \in (S \setminus i)} N(i \cap j) \newline + \frac{1}{3} \sum_{j \in (S \setminus i)} \sum_{k \in (S \setminus i \setminus j)} N(i \cap j \cap k) + ....
$$
where $S$ is the set of all information units $\{VC, C, M, L, VL\}$ and $N(x)$ gives the number of followers of pages in unit(s) $x$.

\vspace{1mm}
\noindent {\bf Information Segregation among Racial Groups} \\
Facebook ad interface returns four racial categories for the users: {\bf Caucasian, African American, Asian American,} and {\bf Hispanic}. For every information unit, we compute the personhoods belonging to each race, and then measure information segregation among them. 
Figure~\ref{fig:segregation}(a) shows the evenness of different racial groups. We can see in Figure~\ref{fig:segregation}(a) that Hispanics have most even access to different political information units; whereas, Asian Americans have most uneven access to political information units.

\vspace{1mm}
\noindent {\bf Information Segregation between Political Groups} \\
Similar to the racial categories, we also computed the personhoods w.r.t. different political leanings for every information unit, 
and then measure the information segregation among these groups. Figure~\ref{fig:segregation}(b) shows that 
conservative leaning users tend to get information evenly from information units; whereas, very conservative leaning users have most uneven access to different units. 
Then to measure how very conservative leaning users have common access to information units with others, 
we plot their joint exposure with other groups in Figure~\ref{fig:segregation}(c). We observe 
that very conservative leaning users have highest joint exposure with conservatives, denoting that they are exposed to multiple information units together. Whereas, they have least joint exposure with very liberal leaning users, implying that these two groups have access to very different information units.

\section{Conclusion}
In this position paper, we proposed five measures of information segregation motivated by the residential segregation measures proposed in literature. 
Then, using these measures, we computed information segregation among US-based Facebook users. Our future work lies in evaluating how search / recommender systems are exposing information to different groups of users, and proposing mechanisms to keep information segregation to acceptable limits.

\vspace{1mm}
\noindent\textbf{Acknowledgments:}
The authors thank the anonymous reviewers whose suggestions helped to improve the paper. 
A. Chakraborty is a recipient of Google India PhD Fellowship and
Prime Minister's Fellowship Scheme for Doctoral Research,
a public-private partnership between Science \& Engineering Research Board (SERB),
Department of Science \& Technology, Government of India and
Confederation of Indian Industry (CII).

\balance

{
\bibliographystyle{ACM-Reference-Format}
\bibliography{Main}
}

\end{document}